\documentclass{article}
\usepackage{spconf,amsmath,graphicx,algorithm,algorithmic,adjustbox,setspace}

\usepackage{color}
\usepackage{xcolor}
\usepackage{enumitem}
\usepackage{hyperref}
\usepackage[capitalize]{cleveref}
\usepackage{multirow}
\usepackage{etoolbox,siunitx}
\robustify\bfseries
\usepackage{booktabs}
\usepackage{balance}
\usepackage{amssymb}
\usepackage{bm}
\usepackage{arydshln}
\usepackage{subcaption}
\usepackage{cite}
\usepackage{hyphenat}
\usepackage{color, colortbl}
	
\definecolor{LightCyan}{rgb}{0.88,1,1}

\DeclareMathOperator*{\argmin}{arg\,min}

\makeatletter
\newcommand{\hathat}[1]{%
\begingroup%
  \let\macc@kerna\z@%
  \let\macc@kernb\z@%
  \let\macc@nucleus\@empty%
  \hat{\raisebox{.35ex}{\vphantom{\ensuremath{#1}}}\smash{\hat{#1}}}%
\endgroup%
}
\makeatother

\let\OLDthebibliography\thebibliography
\renewcommand\thebibliography[1]{
  \OLDthebibliography{#1}
  \setlength{\parskip}{2pt}
  \setlength{\itemsep}{2pt plus 0.4ex}
}
\title{Cold Diffusion for Speech Enhancement}
\name{Hao Yen$^{1,2}$, Fran\c{c}ois G. Germain$^1$, Gordon Wichern$^1$, Jonathan Le Roux$^1$}
\address{$^1$Mitsubishi Electric Research Laboratories (MERL), Cambridge, MA, USA\\
$^2$School of Electrical and Computer Engineering, Georgia Institute of Technology, GA, USA}
\begin{document}
\ninept

\maketitle
\begin{abstract}
Diffusion models have recently shown promising results for difficult enhancement tasks such as the conditional and unconditional restoration of natural images and audio signals. In this work, we explore the possibility of leveraging a recently proposed advanced iterative diffusion model, namely cold diffusion, to recover clean speech signals from noisy signals. The unique mathematical properties of the sampling process from cold diffusion could be utilized to restore high-quality samples from arbitrary degradations. Based on these properties, we propose an improved training algorithm and objective to help the model generalize better during the sampling process. We verify our proposed framework by investigating two model architectures. Experimental results on benchmark speech enhancement dataset VoiceBank-DEMAND demonstrate the strong performance of the proposed approach compared to representative discriminative models and diffusion-based enhancement models. 

\end{abstract}
\section{Introduction}
\label{sec:intro}
Speech enhancement (SE) aims at improving the intelligibility and quality of speech, especially in scenarios where the degradations are caused by non-stationary additive noise. It finds real-world applications in various contexts such as robust automatic speech recognition~\cite{Li2014, Erdogan2015, Chen2015}, speaker recognition~\cite{Vik2019, Taherian2020}, and assistive listening devices~\cite{Wang2017, Healy2019}. Modern state-of-the-art speech enhancement methods based on deep learning, typically estimate a noisy-to-clean mapping through discriminative methods. Time-frequency (T-F) domain methods learn that mapping between spectro-temporal features such as the spectrogram, typically obtained via a short-time Fourier transform (STFT). Some approaches predict the clean speech features directly from the noisy speech features using nonlinear regression techniques, using the clean speech features as training target~\cite{Lu2013, Yong2014}. Others instead predict a T-F mask to estimate the clean speech features through pointwise multiplication between the mask and the noisy speech features~\cite{Erdogan2017, Wang2014}. Time-domain methods learn the noisy-to-clean mapping directly between waveforms, using the clean waveform as training target, in an attempt to circumvent distortions caused by inaccurate phase estimation~\cite{Germain2019, Luo2019}.

Instead of learning a direct noisy-to-clean mapping, a more recent class of approaches uses generative models. Generative models aim to learn the distribution of clean speech as a prior for speech enhancement. Several approaches have utilized deep generative models for speech enhancement using generative adversarial networks (GANs)~\cite{segan, metricgan}, variational autoencoders (VAEs)~\cite{Carbajal2021, Fang2021, Bie2022}, and flow-based models~\cite{Strauss2021}.

The diffusion probabilistic model, proposed in~\cite{Sohl2015}, has shown strong generation and denoising capability in the computer vision field. The standard diffusion probabilistic model includes a diffusion/forward process and a reverse process. The core idea of diffusion process is to gradually convert clean input data to pure noise (isotropic Gaussian distribution), by adding Gaussian noise to the original signal with various steps~\cite{ddpm, ddim}. In the reverse process, the diffusion probabilistic model learns to invert the diffusion process by estimating a noise signal and uses the predicted noise signal to restore the clean signal by subtracting it from the noisy input step by step. Recently, diffusion-based generative models have been introduced to the task of speech enhancement. Lu et al.~\cite{Lu2021} first proposed to build upon standard diffusion framework and devised a supportive reverse process to perform speech enhancement. In their follow-up paper, they further designed a conditional diffusion probabilistic model (CDiffuSE) with a more generalized forward and reverse process which incorporates the noisy spectrograms as the conditioner into the diffusion process~\cite{Lu2022}. In~\cite{welker2022speech} the authors present a complex STFT-based diffusion procedure for speech enhancement, while~\cite{serra2022}, proposes a score-based diffusion model for a universal speech enhancement system that tackles 55 different distortions at the same time.

While existing diffusion models typically built upon additive Gaussian noise for the forward and reverse processes, cold diffusion~\cite{cold2022} considers a broader family of degradation processes (e.g., blur, masking, and downsampling) that can generalize the previous diffusion probabilistic framework without its theoretical limitations. With their proposed improved sampling procedure, cold diffusion shows that the generalization of diffusion models enables us to restore images with arbitrary degradations. The underlying properties of cold diffusion make it a promising framework for speech enhancement where, in realistic conditions, the noise characteristics are usually non-Gaussian. Based on these properties, we expect to be able to avoid the need for any prior assumptions on the noise distribution and recover clean speech signals from arbitrary noise degradations.

In this work, we propose utilizing the cold diffusion framework to perform speech enhancement. Defining the degradation as the deterministic process that iteratively converts clean samples to noisy samples, the model learns to restore clean speech from noisy speech. Furthermore, we propose a modified training process, namely unfolded training, that encourages the network to take into account multiple degradation and restoration steps, thus improving the performance and stability of the restoration model. Experimental results on the VoiceBank-DEMAND dataset demonstrate that our proposed system outperforms existing diffusion-based enhancement models and substantially shrinks the gap typically observed between generative and discriminative models. In summary, the major contributions of the present work are as follows:
(1) this is the first study that investigates the applicability of cold diffusion to additive degradations in general, and SE tasks in particular, with promising results;
(2) we propose an improved training process for cold diffusion to achieve better performance.

\begin{figure}[t]\vspace{-.42cm}
\begin{algorithm}[H]
  \caption{Training for Cold Diffusion}
  \label{alg:cd}
\begin{algorithmic}
  \STATE \textbf{for} $n=1, \dots, N_{\text{iter}}$ \textbf{do} \\
  \STATE \hspace{4mm} Sample clean data $x_0$ \\
  \STATE \hspace{4mm} Sample $t \sim \operatorname{Uniform}(\{1, \dots, T\})$ \\
  \STATE \hspace{4mm} $x_t \leftarrow D(x_0,t)$, $\hat{x}_0 \leftarrow R_\theta(x_t,t)$ \\
  \STATE \hspace{4mm} Take gradient descent step on $\nabla_\theta \left\|\hat{x}_0-x_0\right\|_1$ \\
  \STATE \textbf{end for}
\end{algorithmic}
\end{algorithm}
\vspace{-.7cm}
\end{figure}

\vspace{-.3cm}
\section{Related Methods}
\vspace{-.1cm}

\subsection{Cold Diffusion}
\label{sec:cold}
The original cold diffusion approach \cite{cold2022} is built around two components, a degradation operator $D$ and a restoration operator $R$. Given a ``clean'' training image $x_0\in \mathbb{R}^N$, $D$ is first defined as performing a target degradation of $x_0$ resulting in a ``degraded'' image $y\!=\!D(x_0, T)$. $T$ is a pre-defined number which corresponds to the numbers of severity levels for the degradation, and simultaneously the numbers of \textit{diffusion steps} we will use to reconstruct a clean image from a degraded output. Next, the definition of $D$ is expanded to produce degraded images $x_t$ with an intermediary level of severity $t$ ($0\!\leq\!t\!\leq\!T$) so that $x_t\!=\!D(x_0,t)$. Note that, by definition, $y\!=\!x_T$. A learnable restoration operator $R_\theta$, implemented as a neural network parameterized by $\theta$, is trained to approximately invert $D$, such that $R_\theta(x_t,t)\!\approx\!x_0$. In practice, the training process (the restoration network) is trained via a minimization problem
\begin{equation}
\argmin_\theta \mathbb{E}_{x_0}\left\|R_\theta(D(x_0, t), t)-x_0\right\|,
\end{equation}
where $\|\cdot\|$ denotes a norm, which for audio signals can for example be the $L_1$ norm.
The training process is summarized in {Algorithm}~\ref{alg:cd}. 

After choosing the degradation $D$ and training the model $R_\theta$, these operators can be used in tandem to restore degraded signals whose degradations are similar in nature to the chosen degradation $D$. For small degradations, a \emph{direct reconstruction} consisting of a single reconstruction step $R_\theta(y,T)$ can be used to obtain a restored signal. However, for more severe degradations, direct reconstruction yields poor results. To address this limitation, the cold diffusion approach instead performs an iterative algorithm, applying the restoration operator to a degraded image to perform reconstruction and then (re)degrading the reconstructed image, with the level of degradation severity $t$ decreasing over time, starting from chosen $T$ down to $0$. This iterative method is referred to as \emph{sampled reconstruction}. A further algorithmic improvement is presented in~\cite{cold2022} where the sampling is modified by altering the naive (re)degradation step in the iteration with a first-order approximation of the degradation operator $D$ as shown here in {Algorithm}~\ref{alg:sampling}. The improvement is shown to result in a reconstruction process that is then much more tolerant to errors in the estimation of $R_\theta$.

\vspace{-.3cm}
\subsection{Conditional Diffusion Probabilistic Model}
\vspace{-.1cm}
\label{sec:cdiffuse}
The conditional diffusion probabilistic model for speech enhancement (CDiffuSE)~\cite{Lu2022} is a generalized version of the prior diffusion probabilistic model for speech enhancement (DiffuSE)~\cite{Lu2021}, which was the first study to apply this type of model to SE tasks. DiffuSE did not take into account the noisy data but used Gaussian noise solely during the diffusion/reverse process, which is not a valid assumption under realistic conditions. To address this issue, CDiffuSE defines the conditional diffusion process by incorporating the noisy data into the diffusion process and assumes that the mean of the Markov chain Gaussian model of a given step is represented as a linear interpolation between the clean data and noisy data. Under this assumption, the model learns to estimate both the Gaussian noise and the non-Gaussian noise during the reverse process. The authors derive the corresponding optimization criterion for the conditional diffusion and reverse processes, and show that the resulting model is a generalization of the original diffusion probabilistic model.

\begin{figure}[t]\vspace{-.42cm}
\begin{algorithm}[H]
  \caption{Improved Sampling for Cold Diffusion \cite{cold2022}}
  \label{alg:sampling}
\begin{algorithmic}
  \STATE \textbf{Input:} A degraded sample $x_T$ \\
  \STATE \textbf{for} $t=T,T-1,\dots, 1$ \textbf{do} \\
  \STATE \hspace{4mm} $\hat{x}_0 \leftarrow R_\theta(x_t,t)$ \\
  \STATE \hspace{4mm} $x_{t-1} \leftarrow x_t-D(\hat{x}_0,t)+D(\hat{x}_0,t-1)$ \\
  \STATE \textbf{end for}
\end{algorithmic}
\end{algorithm}
\vspace{-.6cm}
\end{figure}

However, despite conditioning on noisy spectrograms, the derivation of the CDiffuSE objective function is still based on the assumption that the distribution of the noisy speech follows a standard white Gaussian, which may not be the case for speech enhancement as described in the following section.
\vspace{-.2cm}
\section{Cold Diffusion for Speech Enhancement}
\label{sec:propose}

\subsection{Degradation and Sampling Process}
\label{sec:deg}
We propose to formulate the speech enhancement problem, which is to recover clean speech $x_0$ from noisy speech $y=x_0+n$, within the cold diffusion framework. We do so by defining a degradation process along the lines of the \textit{animorphosis} transformation in~\cite{cold2022}, where a ``clean'' sample (image of a person) is iteratively transformed into an out-of-domain ``degraded'' sample (picture of an animal). However, note that our process differs in the sense that our degraded sample still contains the clean sample information, as we now have a degradation process that instead \textit{adds} an out-of-domain sample to the clean sample. Such a degradation does not correspond to any of those addressed in \cite{cold2022}.
More formally, given a clean sample $x_0$ and the noisy data $x_T=y=x_0+n$, we define the degraded sample for a level of degradation severity $t$ as 
\begin{equation}
x_t=D_{x_T}(x_0,t)=\sqrt{\alpha_t} x_0+\sqrt{1-\alpha_t} x_T,
\label{eqn:deg}
\end{equation} 
where, departing \cite{cold2022}, we make the dependence on $x_T$ explicit for clarity. 
The degraded sample $x_t$ is the deterministic interpolation between $x_0$ and $x_T$ with interpolation weights defined by $\alpha_t$, with $\alpha_t$ starting from $\alpha_0\!=\!1$ and gradually decreased to $\alpha_T\!=\!0$, where $T$ is the total number of degradation steps (or equivalently the terminal level of degradation severity).

The sampling process follows the improved sampling algorithm from \cite{cold2022}. Given the degraded sample $x_t$ at level $t$, we obtain the restored sample $\hat{x}_0$ from the restoration model $R_\theta$. One possibility for obtaining $x_{t-1}$ would be to use $D(\hat{x}_0,s)=D_{x_T}(\hat{x}_0,s)$ %
in {Algorithm}~\ref{alg:sampling}.
Another possibility, which was found to work better in \cite{cold2022} and is akin to the deterministic sampling in denoising diffusion implicit models \cite{ddim}, is to use an alternative degradation anchored around $x_t$, that is, the degradation which leads from $\hat{x}_0$ to $x_t$ in $t$ steps.
This can be done by defining a modified ``noisy'' sample $\hat{x}_T^{(t)}$ as
\begin{equation}
\hat{x}_T^{(t)}=\frac{1}{\sqrt{1-\alpha_t}}(x_t-\sqrt{\alpha_t} \hat{x}_0),
\end{equation} 
which when used in (\ref{eqn:deg}) in place of $x_T$ leads to a degradation operator
\begin{equation}
s \mapsto D{\raisebox{-1pt}{$\scriptstyle\hat{x}_T^{(t)}$}}\left(\hat{x}_0, s\right) 
=\sqrt{\alpha_{s}} \hat{x}_0+\frac{\sqrt{1-\alpha_{s}}}{\sqrt{1-\alpha_t}} (x_t -\sqrt{\alpha_t}\hat{x}_0)
\label{eqn:Dupdate}
\end{equation}
that verifies $D{\raisebox{-1pt}{$\scriptstyle\hat{x}_T^{(t)}$}}(x_0,t)=x_t$.

While it might be counterintuitive at first that the implied degraded sample shifts during the sampling process, this must be understood as an expedient intermediary mathematical quantity from the perspective of a local approximation of the ambiguously-defined $D(\hat{x}_0,t)$ and $D(\hat{x}_0, t{-}1)$ rather than to be interpreted literally as our initial degraded output being changed. The calculation of $x_{t-1}$ in Algorithm~\ref{alg:sampling} then simplifies to
\begin{equation}
x_{t-1} \leftarrow \sqrt{\alpha_{t-1}} \hat{x}_0+\frac{\sqrt{1-\alpha_{t-1}}}{\sqrt{1-\alpha_t}} (x_t -\sqrt{\alpha_t}\hat{x}_0).
\end{equation}

Additionally, we show in Section \ref{sec:exp} that this formulation gets better performance than the alternative proposition mentioned earlier, where we simply use $D_{x_T}(\hat{x}_0,s)$.

\begin{figure}[t]\vspace{-.42cm}
\begin{algorithm}[H]
  \caption{Proposed Unfolded Training for Cold Diffusion}
  \label{alg:mcd}
\begin{algorithmic}
  \STATE \textbf{for} $n=1, ..., N_{\text{iter}}$ \textbf{do} \\
  \STATE \hspace{4mm} Sample clean data $x_0$ \\
  \STATE \hspace{4mm} Sample $t \sim \operatorname{Uniform}(\{1, \dots, T\})$ \\
  \STATE \hspace{4mm} $x_t \leftarrow D(x_0,t)$, $\hat{x}_0 \leftarrow R_\theta(x_t,t)$ \\
  \STATE \hspace{4mm} Sample $t' \sim \operatorname{Uniform}(\{1, \dots, t\})$ \\
  \STATE \hspace{4mm} $\hat{x}_{t'} \leftarrow D(\hat{x}_0,t')$, $\hathat{x}_0 \leftarrow R_\theta(\hat{x}_{t'},t')$ \\
  \STATE \hspace{4mm} Take gradient descent step on $\nabla_\theta ( \left\|\hat{x}_0-x_0\right\|_1 +  \left\|\hathat{x}_0-x_0 \right\|_1)$ \\
  \STATE \textbf{end for}
\end{algorithmic}
\end{algorithm}
\vspace{-.8cm}
\end{figure}

\subsection{Unfolded Training for Cold Diffusion}

While our proposed cold diffusion-based speech enhancement network can be trained similarly to the original cold diffusion method, we find that the original cold diffusion training procedure \cite{cold2022} suffers from limitations. As can be seen in Algorithm~\ref{alg:cd}, the network only gets to see degradations resulting from the forward diffusion process and attempts to compensate for those, but it has no way to compensate for errors in its attempt at reconstructing the clean input. We thus propose an unfolded training approach that allows the network to consider and potentially repair its own past mistakes.

We propose to improve the training algorithm by unfolding multiple degradation and restoration steps, using two steps in the following as a proof of concept. As in the original cold diffusion training process, we first transform a clean sample $x_0$ to its degraded version $x_t$ with respect to severity $t$ using the degradation operator $D$, then apply the restoration operator $R_\theta$ to obtain a first predicted clean sample $\hat{x}_0$. We then generate another degraded sample $\hat{x}_{t'}$. However, instead of using another clean sample, we use the predicted clean sample $\hat{x}_0$ from the last step and perform degradation with a smaller severity $t' \leq t$. We then restore $\hat{x}_{t'}$ to another approximated clean sample $\hathat{x}_0$. As shown in {Algorithm}~\ref{alg:mcd}, the unfolded training objective is now defined to reduce the $L_1$ distance between each of the estimated samples $\hat{x}_0$ and $\hathat{x}_0$ and the clean sample $x_0$:
\begin{align}
    \mathcal{L}_{\text{unf}}(\theta) &= \| \hat{x}_0-x_0 \|_1 + \| \hathat{x}_0-x_0 \|_1 \nonumber \\
    &\hspace{-.4cm}=\|R_\theta(D(x_0, t), t)\!-\!x_0\|_1 +  \|R_\theta(D(\hat{x}_0, t'), t')\!-\!x_0\|_1.
\end{align}
using Eq.~\ref{eqn:Dupdate} for $D(\hat{x}_0, t')$. We argue that the combination of unfolded steps is more consistent with the iterative sampling process of cold diffusion, making the model more tolerant of errors in $R_\theta$.

\subsection{Model Structure}
\label{sec:model}
We consider two different backbone network architectures.

\begin{figure}[t]
     \centering
     \begin{subfigure}[t]{\columnwidth}
         \centering
         \includegraphics[width=\linewidth]{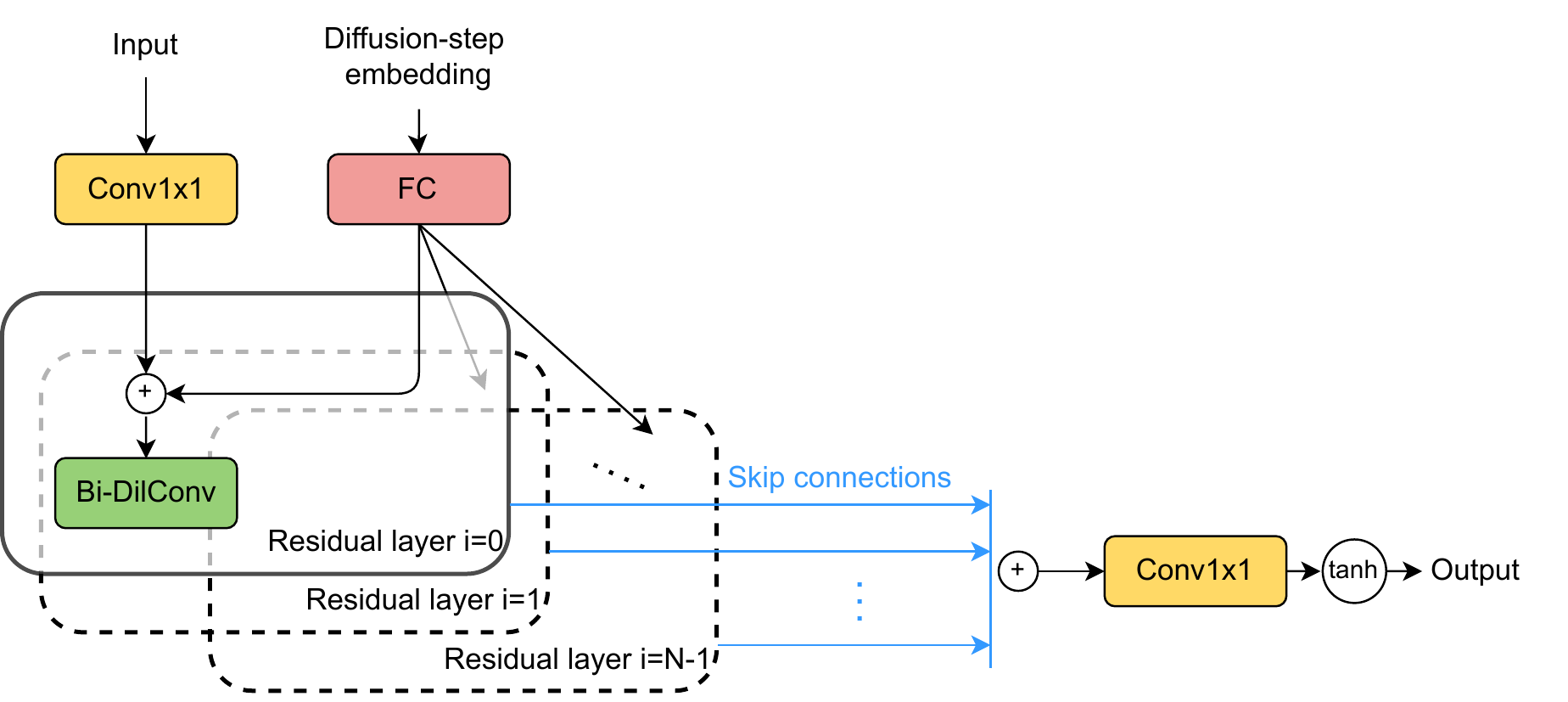}
         \caption{Architecture of the DiffWave model.}
         \label{fig:diffwave}
     \end{subfigure}
     \hfill
     \begin{subfigure}[t]{\columnwidth}
         \centering
         \includegraphics[width=\linewidth]{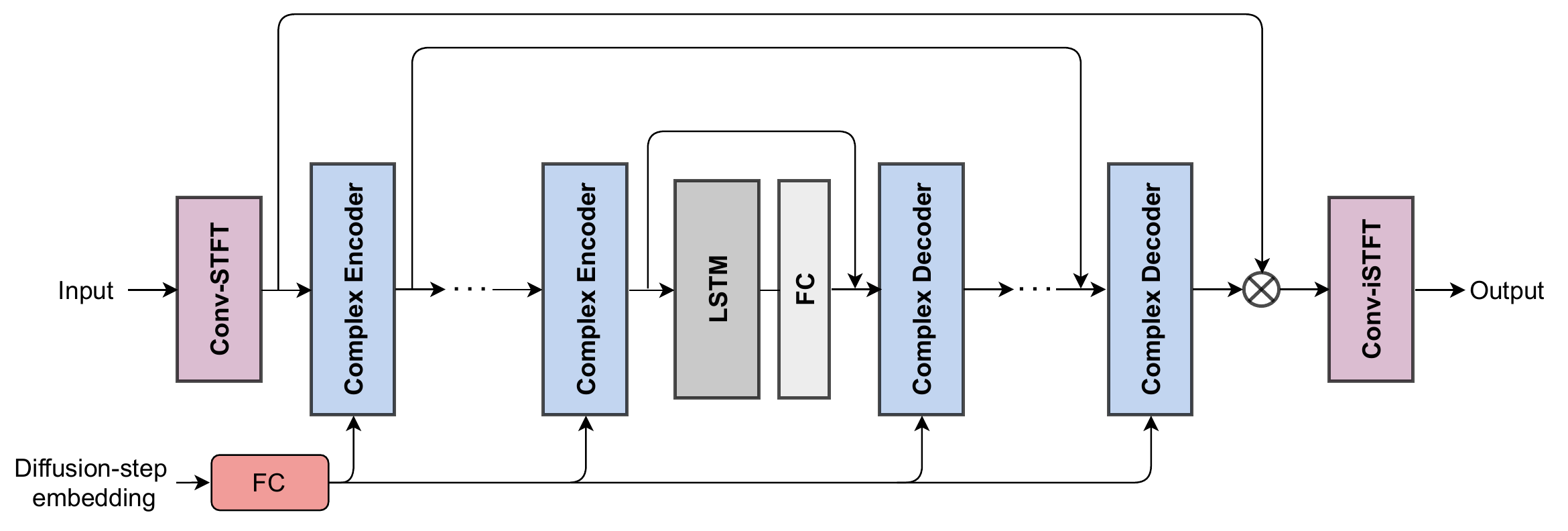}
         \caption{Architecture of the DCCRN model.}
         \label{fig:dccrn}
     \end{subfigure}
     \vspace{-.2cm}
        \caption{Architectures of the backbone models used in this work. FC denotes a fully connected network.}
             \vspace{-.3cm}
        \label{fig:model}
\end{figure}

\noindent \textbf{DiffWave:} The DiffWave~\cite{Kong2021} model architecture is similar to WaveNet~\cite{Oord2016}. DiffWave uses a feed-forward and bidirectional dilated convolution (Bi-DilConv) architecture, which is non\hyp{}autoregressive and can synthesize high-dimensional waveforms in parallel. The network is composed of a stack of $N$ residual layers with residual channels $C$. These layers are grouped into $m$ blocks and each block has $n=\frac{N}{m}$ layers. The dilation is doubled at each layer within each block, i.e., $[1,2,4,\dots,2^{n-1}]$. The skip connections from all residual layers are summed up as in WaveNet. The original DiffWave uses a ReLU activation function before the output. However, unlike the original DiffWave, which aims to estimate noise at each step, our system directly estimates the clean waveform. Hence, we modify the last activation from ReLU to Tanh, directly generating an output waveform. Figure~\ref{fig:diffwave} shows the overall architecture.

\noindent \textbf{DCCRN:} Deep Complex Convolution Recurrent Network (DCCRN)~\cite{Hu2020} modified the original CRN~\cite{Tan2018} with a complex CNN and complex batch normalization layers in the encoder and decoder. Specifically, the complex module models the correlation between magnitude and phase with the simulation of complex multiplication. When training, DCCRN estimates a complex ratio mask (CRM)~\cite{Williamson2016} and is optimized by waveform approximation (WA) on the reconstructed signal. The complex encoder block includes complex Conv2d, complex batch normalization~\cite{Trabelsi2017}, and real-valued PReLU~\cite{He2015}. Complex Conv2d consists of four traditional Conv2d operations, controlling the complex information flow throughout the encoder. We adapt DCCRN by inserting a \textit{diffusion-step embedding layer} into all encoder/decoder blocks, providing the model with information of the diffusion (degradation) step $t$. The {diffusion-step embedding layer} uses a sinusoidal positional embedding followed by a fully connected layer. Fig.~\ref{fig:dccrn} shows the overall architecture.

\vspace{-.1cm}
\section{Experiments}
\label{sec:exp}
\vspace{-.1cm}

\subsection{Dataset}
To train and evaluate our model, following CDiffuSE~\cite{Lu2022}, we use the VoiceBank-DEMAND dataset~\cite{Voicebank2017} spoken by 30 speakers with 10 types of noises. The dataset is split into a training and a testing set with 28 and 2 speakers. Four types of signal-to-noise ratio (SNRs) are used to mix clean samples with noise samples in the dataset, $[0,5,10,15]$ dB for training and $[2.5,7.5, 12.5, 17.5]$ dB for testing. We further excerpt two speakers from the training set to form the validation set, resulting in 10,802 utterances for training and 770 for validation. The testing set has 824 utterances.

We follow CDiffuSE~\cite{Lu2022} and use multiple evaluation measurements, including wide-band perceptual evaluation of speech quality (PESQ)~\cite{pesq}, prediction of the signal distortion (CSIG), prediction of the background intrusiveness (CBAK), and prediction of the overall speech quality (COVL)~\cite{Hu2008}. More specifically, PESQ assesses the perceptual quality of speech signals, and CSIG, CBAK, and COVL are composite metrics reflecting mean opinion scores (MOS).

\vspace{-.05cm}
\subsection{Model Setting and Training Procedure}
\vspace{-.05cm}
We investigate two model architectures, DiffWave and DCCRN (see Section~\ref{sec:model}). For DiffWave, we broadly follow the setup of DiffuSE and CDiffuSE. We construct the model using 30 residual layers with 3 dilation cycles and a kernel size of 3. While DiffuSE and CDiffuSE use the DiffWave version with mel-filterbank conditioner (pretrained for the former, not for the latter), ours is the unconditioned version (cf.\ Fig.~\ref{fig:diffwave}), resulting in a slightly smaller model with 2.3M parameters overall. %
For DCCRN, the number of channels in encoder/decoder is $\{32,64,128,128,256,256\}$ and the kernel size and stride are set to $(5,2)$. The adapted DCCRN with diffusion-step embedding layer has around 5.6M trainable parameters. Our systems take $T\!=\!50$ diffusion steps. The interpolation parameter $\alpha_t$ is defined using a cosine schedule as proposed in~\cite{Nichol2021}. More formally,\vspace{-.125cm}
\begin{equation}
\alpha_t=\frac{f(t)}{f(0)},\quad f(t)=\cos \left(\frac{t / T+s}{1+s} \cdot \frac{\pi}{2}\right)^2, \quad s=0.008,
\vspace{-.075cm}\end{equation} 
which satisfies $\alpha_0\!=\!1$ and $\alpha_T\!=\!0$. We train our model with $N_\text{iter}\!=\!10^5$ iterations and choose the best model using the PESQ score on the validation set. We use $L_1$ loss over all output samples in a batch and set the batch size to 256. For our proposed cold diffusion-based method, we report the results for both direct reconstruction (1 step) and improved sampling (50 steps) as mentioned in Section~\ref{sec:cold}. %

\begin{table}[t!]
\caption{Comparison of various (discriminative models, DiffuSE, CDiffuSE) and our proposed cold diffusion-based methods on VoiceBank-DEMAND. CD refers to cold diffusion with the original training, and Unfolded CD denotes cold diffusion with our proposed unfolded training. ``w/ $D_{x_T}$'' indicates that $D_{x_T}(\hat{x}_0, s)$ is used for the degradation (cf.~Section~\ref{sec:deg}). * indicates results reported as-is from prior literature.}\vspace{-.35cm}
\label{tab:1}
\setlength\tabcolsep{3.25pt}
\begin{footnotesize}
\begin{adjustbox}{width=\columnwidth}
\begin{tabular}{lcccccc}
\toprule
\textbf{Method}                     & \textbf{Network}          & \textbf{Steps} & \textbf{PESQ} & \textbf{CSIG} & \textbf{CBAK} & \textbf{COVL} \\ \midrule \midrule
Unprocessed                         & --                        & --             & 1.97          & 3.37          & 2.45          & 2.65          \\ \midrule
Conv-TasNet*~\cite{Luo2019,Lu2022}  & --                        & --             & 2.84          & 2.33          & 2.62          & 2.51          \\
DiffWave (uncond.)~\cite{Kong2021}  & --                        & --             & 2.49          & 3.67          & 3.27          & 3.07          \\
DiffWave (cond.)~\cite{Kong2021}    & --                        & --             & 2.52          & 3.72          & 3.27          & 3.11          \\
DCCRN~\cite{Hu2020,Lv2022}          & --                        & --            & 2.59          & 3.71          & 3.23          & 3.13          \\
WaveCRN*~\cite{Hsieh2020}            & --                        & --             & 2.64          & 3.94          & 3.37          & 3.29          \\
Demucs~\cite{Dfossez2020}           & --                        & --             & \textbf{3.07} & \textbf{4.31} & \textbf{3.40} & \textbf{3.63} \\ \midrule \midrule
DiffuSE (Base)*~\cite{Lu2021}        & \multirow{4}{*}{DiffWave} & 50             & 2.41          & 3.61          & 2.81          & 2.99          \\
DiffuSE (Large)*~\cite{Lu2021}       &                           & 200            & 2.43          & 3.63          & 2.81          & 3.01          \\
CDiffuSE (Base)*~\cite{Lu2022}       &                           & 50             & 2.44          & 3.66          & 2.83          & 3.03          \\
CDiffuSE (Large)*~\cite{Lu2022}      &                           & 200            & 2.52          & 3.72          & 2.91          & 3.10          \\ \midrule
CD                                  & \multirow{5}{*}{DiffWave} & 1              & 2.42          & 3.53          & 3.15          & 2.97          \\
CD                                  &                           & 50             & 2.48          & 3.75          & 3.02          & 2.97          \\
Unfolded CD                         &                           & 1              & 2.50          & 3.59          & 3.21          & 3.04          \\
Unfolded CD                         &                           & 50             & 2.60          & 3.79          & 3.21          & 3.19          \\ 
Unfolded CD w/ $D_{x_T}$            &                           & 50             & 2.55          & 3.69          & 3.18          & 3.10         \\
\midrule
CD                                  & \multirow{3}{*}{DCCRN}    & 50             & 2.69          & 3.83          & 3.28          & 3.27          \\
Unfolded CD                         &                           & 50             & \textbf{2.77} & \textbf{3.91} & \textbf{3.32} & \textbf{3.33}  \\ 
Unfolded CD w/ $D_{x_T}$            &                           & 50             & 2.68          & 3.80          & 3.25          & 3.23   \\
\bottomrule
\end{tabular}
\end{adjustbox}
\end{footnotesize}
 \vspace{-.45cm}
\end{table}

\vspace{-.05cm}
\subsection{Results}
\vspace{-.05cm}

Table~\ref{tab:1} reports the results of representative discriminative models, diffusion-based enhancement models, and our cold diffusion-based methods. For Demucs, we rerun the publicly-available pretrained model. Base and Large DiffuSE/CDiffuSE use 50 and 200 diffusion steps, respectively. For cold diffusion-based methods, we report the results for both the DiffWave and DCCRN architectures. We also retrain and report results of discriminatively-trained DiffWave (conditioned and unconditioned) and DCCRN models for fair comparison. The original training framework for cold diffusion is denoted as CD. Our proposed unfolded training framework is denoted as Unfolded CD. Except where indicated, all cold diffusion-based models use $D{\raisebox{-1pt}{$\scriptstyle\hat{x}_T^{(t)}$}}(\hat{x}_0, s)$ as degradation operator (cf.\ Section~\ref{sec:deg}). Table~\ref{tab:1} shows using $D_{x_T}(\hat{x}_0, s)$ leads to slightly worse performance.

Comparing CD and Unfolded CD with two diffusion-based methods, DiffuSE~\cite{Lu2021} and CDiffuSE~\cite{Lu2022}, we find an improvement with the same DiffWave model as backbone architecture. CD outperforms DiffuSE as well as Base CDiffuSE on all evaluation metrics, and unfolded CD yields further improvements and outperforms Large CDiffuSE with fewer sampling steps and no conditioning mechanism. 
Comparing the results of our CD framework combined with two different backbone models, we find that both CD and unfolded CD with the DCCRN model outperform the DiffWave model. This shows that the cold diffusion framework significantly benefits from increased model capacity. Moreover, unfolded CD on both models shows significant improvement over CD on all the metrics. 
Additionally, comparing the best results we obtained with our proposed framework to existing discriminative models, we see that, while they do not yet compete with top-performing methods, we make up much of the ground that exists between them and the best results of prior diffusion-based methods, namely Large CDiffuSE. We note however that some of those methods benefit from far higher model capacity than the backbone models we used, and that their performance on the VoiceBank-DEMAND dataset has been shown to significantly benefit from techniques such as data augmentation \cite{Dfossez2020}, whose inclusion we leave to future work. 
Most importantly, except for the CBAK score for DiffWave, each Unfolded CD model improves upon the discriminative model with the same backbone network (DiffWave or DCCRN) on all the metrics.

\vspace{-.1cm}
\section{Conclusion}
\vspace{-.1cm}
In this study, we proposed to use cold diffusion for speech enhancement. To further improve the framework, we also proposed an unfolded training process that allows the model to learn from multiple degradation and restoration steps. Our results show that the cold diffusion framework can yield better performance than other diffusion-based enhancement models and our proposed unfolded training effectively improves the original framework. While our systems have yet to achieve the best overall results, we significantly shrink the performance gap between diffusion-based models and discriminative models. We contend that the remaining gap can be closed with different backbone models, advanced training losses, and data augmentation, all of which are compatible with our framework, and consider those to be important directions for future work. Also, our paper focused on establishing the in-domain performance of cold diffusion, but we take note that CDiffuSE also displayed strong out-of-domain performance (i.e., models trained on VoiceBank\hyp{}\-DEMAND worked well on other datasets). We contend that this robustness is due to its conditioning on the noisy input, and consider such an addition to our framework to be another direction for future work.

\balance 
\footnotesize
\bibliographystyle{IEEEtran}
\bibliography{refs}

\begin{thebibliography}{10}
\providecommand{\url}[1]{#1}
\csname url@samestyle\endcsname
\providecommand{\newblock}{\relax}
\providecommand{\bibinfo}[2]{#2}
\providecommand{\BIBentrySTDinterwordspacing}{\spaceskip=0pt\relax}
\providecommand{\BIBentryALTinterwordstretchfactor}{4}
\providecommand{\BIBentryALTinterwordspacing}{\spaceskip=\fontdimen2\font plus
\BIBentryALTinterwordstretchfactor\fontdimen3\font minus
  \fontdimen4\font\relax}
\providecommand{\BIBforeignlanguage}[2]{{%
\expandafter\ifx\csname l@#1\endcsname\relax
\typeout{** WARNING: IEEEtran.bst: No hyphenation pattern has been}%
\typeout{** loaded for the language `#1'. Using the pattern for}%
\typeout{** the default language instead.}%
\else
\language=\csname l@#1\endcsname
\fi
#2}}
\providecommand{\BIBdecl}{\relax}
\BIBdecl

\bibitem{Li2014}
J.~Li, L.~Deng, Y.~Gong, and R.~Haeb-Umbach, ``An overview of noise-robust
  automatic speech recognition,'' \emph{IEEE/ACM Trans. Audio, Speech, Lang.
  Process.}, vol.~14, no.~4, pp. 745--777, 2014.

\bibitem{Erdogan2015}
H.~Erdogan, J.~R. Hershey, S.~Watanabe, and J.~Le~Roux, ``Phase-sensitive and
  recognition-boosted speech separation using deep recurrent neural networks,''
  in \emph{Proc. ICASSP}, 2015.

\bibitem{Chen2015}
Z.~Chen, S.~Watanabe, H.~Erdogen, and J.~R. Hershey, ``Speech enhancement and
  recognition using multi-task learning of long short-term memory recurrent
  neural networks,'' in \emph{Proc. Interspeech}, 2015.

\bibitem{Vik2019}
M.~L. Vik, ``Speech enhancement with a generative adversarial network,''
  Master's thesis, NTNU, Trondheim, Norway, 2019.

\bibitem{Taherian2020}
H.~Taherian, Z.-Q. Wang, J.~Chang, and D.~Wang, ``Robust speaker recognition
  based on single-channel and multi-channel speech enhancement,''
  \emph{IEEE/ACM Trans. Audio, Speech, Lang. Process.}, vol.~28, pp.
  1293--1302, 2020.

\bibitem{Wang2017}
D.~Wang, ``Deep learning reinvents the hearing aid,'' \emph{IEEE Spectr.},
  vol.~54, no.~03, pp. 32--37, 2017.

\bibitem{Healy2019}
E.~Healy, J.~Vasko, and D.~Wang, ``The optimal threshold for removing noise
  from speech is similar across normal and impaired hearing—a time-frequency
  masking study,'' \emph{J. Acoust. Soc. Am.}, vol. 145, no.~06, pp.
  EL581--EL586, 2019.

\bibitem{Lu2013}
X.~Lu, Y.~Tsao, S.~Matsuda, and C.~Hori, ``Speech enhancement based on deep
  denoising autoencoder,'' in \emph{Proc. Interspeech}, 2013.

\bibitem{Yong2014}
Y.~Xu, J.~Du, L.-R. Dai, and C.-H. Lee, ``An experimental study on speech
  enhancement based on deep neural networks,'' \emph{IEEE Signal Process.
  Lett.}, vol.~21, no.~1, pp. 65--68, 2014.

\bibitem{Erdogan2017}
H.~Erdogan, J.~R. Hershey, S.~Watanabe, and J.~Le~Roux, ``Deep recurrent
  networks for separation and recognition of single-channel speech in
  nonstationary background audio,'' in \emph{New Era for Robust Speech
  Recognition, Exploiting Deep Learning}.\hskip 1em plus 0.5em minus
  0.4em\relax Cham: Springer, 2017, ch.~7, pp. 165--186.

\bibitem{Wang2014}
Y.~Wang, A.~Narayanan, and D.~Wang, ``On training targets for supervised speech
  separation,'' \emph{IEEE/ACM Trans. Audio, Speech, Lang. Process.}, vol.~22,
  no.~12, pp. 1849--1858, 2014.

\bibitem{Germain2019}
F.~G. Germain, Q.~Chen, and V.~Koltun, ``Speech denoising with deep feature
  losses,'' in \emph{Proc. Interspeech}, 2019.

\bibitem{Luo2019}
Y.~Luo and N.~Mesgarani, ``{Conv-TasNet}: Surpassing ideal time-frequency
  magnitude masking for speech separation,'' \emph{IEEE/ACM Trans. Audio,
  Speech, Lang. Process.}, vol.~27, no.~8, pp. 1256--1266, 2019.

\bibitem{segan}
S.~Pascual, A.~Bonafonte, and J.~Serr{\`a}, ``{SEGAN}: Speech enhancement
  generative adversarial network,'' in \emph{Proc. Interspeech}, 2017.

\bibitem{metricgan}
S.-W. Fu, C.-F. Liao, Y.~Tsao, and S.-D. Lin, ``{MetricGAN}: Generative
  adversarial networks based black-box metric scores optimization for speech
  enhancement,'' in \emph{Proc. ICML}, 2019.

\bibitem{Carbajal2021}
G.~Carbajal, J.~Richter, and T.~Gerkmann, ``Disentanglement learning for
  variational autoencoders applied to audio-visual speech enhancement,'' in
  \emph{Proc. WASPAA}, 2021.

\bibitem{Fang2021}
H.~Fang, G.~Carbajal, S.~Wermter, and T.~Gerkmann, ``Variational autoencoder
  for speech enhancement with a noise-aware encoder,'' in \emph{Proc. ICASSP},
  2021.

\bibitem{Bie2022}
X.~Bie, S.~Leglaive, X.~Alameda-Pineda, and L.~Girin, ``Unsupervised speech
  enhancement using dynamical variational autoencoders,'' \emph{IEEE/ACM Trans.
  Audio, Speech, Lang. Process.}, vol.~30, pp. 2993--3007, 2022.

\bibitem{Strauss2021}
M.~Strauss and B.~Edler, ``A flow-based neural network for time domain speech
  enhancement,'' in \emph{Proc. ICASSP}, 2021.

\bibitem{Sohl2015}
J.~Sohl-Dickstein, E.~Weiss, N.~Maheswaranathan, and S.~Ganguli, ``Deep
  unsupervised learning using nonequilibrium thermodynamics,'' in \emph{Proc.
  ICML}, 2015.

\bibitem{ddpm}
J.~Ho, A.~Jain, and P.~Abbeel, ``Denoising diffusion probabilistic models,'' in
  \emph{Proc. NeurIPS}, 2020.

\bibitem{ddim}
J.~Song, C.~Meng, and S.~Ermon, ``Denoising diffusion implicit models,'' in
  \emph{Proc. ICLR}, 2021.

\bibitem{Lu2021}
Y.-J. Lu, Y.~Tsao, and S.~Watanabe, ``A study on speech enhancement based on
  diffusion probabilistic model,'' in \emph{Proc. APSIPA ASC}, 2021.

\bibitem{Lu2022}
Y.-J. Lu, Z.-Q. Wang, S.~Watanabe, A.~Richard, C.~Yu, and Y.~Tsao,
  ``Conditional diffusion probabilistic model for speech enhancement,'' in
  \emph{Proc. ICASSP}, 2022.

\bibitem{welker2022speech}
S.~Welker, J.~Richter, and T.~Gerkmann, ``Speech enhancement with score-based
  generative models in the complex {STFT} domain,'' in \emph{Proc.
  Interspeech}, 2022.

\bibitem{serra2022}
J.~Serr\`{a}, S.~Pascual, J.~Pons, R.~O. Araz, and D.~Scaini, ``Universal
  speech enhancement with score-based diffusion,'' \emph{arXiv preprint
  arXiv:2206.03065}, 2022.

\bibitem{cold2022}
A.~Bansal, E.~Borgnia, H.-M. Chu, J.~S. Li, H.~Kazemi, F.~Huang \emph{et~al.},
  ``Cold diffusion: Inverting arbitrary image transforms without noise,''
  \emph{arXiv preprint arXiv:2208.09392}, 2022.

\bibitem{Kong2021}
Z.~Kong, W.~Ping, J.~Huang, K.~Zhao, and B.~Catanzaro, ``{DiffWave}: A
  versatile diffusion model for audio synthesis,'' in \emph{Proc. ICLR}, 2021.

\bibitem{Oord2016}
A.~van~den Oord, S.~Dieleman, H.~Zen, K.~Simonyan, O.~Vinyals, A.~Graves
  \emph{et~al.}, ``{W}ave{N}et: A generative model for raw audio,'' in
  \emph{Proc. SSW}, 2016.

\bibitem{Hu2020}
Y.~Hu, Y.~Liu, S.~Lv, M.~Xing, S.~Zhang, Y.~Fu \emph{et~al.}, ``{DCCRN}: Deep
  complex convolution recurrent network for phase-aware speech enhancement,''
  in \emph{Proc. Interspeech}, 2020.

\bibitem{Tan2018}
K.~Tan and D.~Wang, ``A convolutional recurrent neural network for real-time
  speech enhancement,'' in \emph{Proc. Interspeech}, 2018.

\bibitem{Williamson2016}
D.~S. Williamson, Y.~Wang, and D.~Wang, ``Complex ratio masking for monaural
  speech separation,'' \emph{IEEE/ACM Trans. Audio, Speech, Lang. Process.},
  vol.~24, no.~3, pp. 483--492, 2016.

\bibitem{Trabelsi2017}
C.~Trabelsi, O.~Bilaniuk, Y.~Zhang, D.~Serdyuk, S.~Subramanian, J.~F. Santos
  \emph{et~al.}, ``Deep complex networks,'' in \emph{Proc. ICLR}, 2018.

\bibitem{He2015}
K.~He, X.~Zhang, S.~Ren, and J.~Sun, ``Delving deep into rectifiers: Surpassing
  human-level performance on {I}mage{N}et classification,'' in \emph{Proc.
  ICCV}, 2015.

\bibitem{Voicebank2017}
C.~Valentini-Botinhao, ``Noisy speech database for training speech enhancement
  algorithms and {TTS} models,'' 2017, \url{https://doi.org/10.7488/ds/2117}.

\bibitem{pesq}
A.~W. Rix, J.~G. Beerends, M.~P. Hollier, and A.~P. Hekstra, ``Perceptual
  evaluation of speech quality ({PESQ})-a new method for speech quality
  assessment of telephone networks and codecs,'' in \emph{Proc. ICASSP}, 2001.

\bibitem{Hu2008}
Y.~Hu and P.~C. Loizou, ``Evaluation of objective quality measures for speech
  enhancement,'' \emph{IEEE Trans. Audio, Speech, Lang. Process.}, vol.~16,
  no.~1, pp. 229--238, 2008.

\bibitem{Nichol2021}
A.~Nichol and P.~Dhariwal, ``Improved denoising diffusion probabilistic
  models,'' in \emph{Proc. ICML}, 2021.

\bibitem{Lv2022}
S.~Lv, Y.~Fu, M.~Xing, J.~Sun, L.~Xie, J.~Huang \emph{et~al.}, ``{S-DCCRN}:
  Super wide band {DCCRN} with learnable complex feature for speech
  enhancement,'' in \emph{Proc. ICASSP}, 2022.

\bibitem{Hsieh2020}
T.-A. Hsieh, H.-M. Wang, X.~Lu, and Y.~Tsao, ``{WaveCRN}: An efficient
  convolutional recurrent neural network for end-to-end speech enhancement,''
  \emph{IEEE Signal Process. Lett.}, vol.~27, pp. 2149--2153, 2020.

\bibitem{Dfossez2020}
A.~D{\'e}fossez, G.~Synnaeve, and Y.~Adi, ``Real time speech enhancement in the
  waveform domain,'' in \emph{Proc. Interspeech}, 2020.

\end{thebibliography}

\end{document}